# Two-photon sweeping out of the *K*-shell of a heavy atomic ion

A.N. Hopersky, A.M. Nadolinsky, S.A. Novikov, *and* R.V. Koneev

Rostov State University of Railway Engineering, 344038, Rostov-on-Don, Russia

Within the framework of the second-order non-relativistic quantum perturbation theory, the methods of the theory of irreducible tensor operators, and nonorthogonal orbitals, the absolute values and shape of the generalized cross-sections of the two-photon double ionization (sweeping) of the *K*-shell of a heavy neon-like ion of an iron atom were predicted. The complete wave functions of the ground state of the ion and the states of its ionization are obtained in the single-configuration Hartree-Fock approximation. The effects of radial relaxation of ionization states in the field of one and two vacancies in the *K*-shell are taken into account. The matrix element of the operator of the radiation transition between continuum-spectrum states is taken into account by introducing a correlation function. It was established that the generalized cross-section of the two-photon sweeping of the *K*-shell at high ($\geq 15.8$ keV) energies of absorbed X-ray photons is several orders of magnitude greater than the generalized cross-section of the two-photon single ionization of the *K*-shell. This result qualitatively reproduces the work of Novikov S.A. and Hopersky A.N. (J. Phys. B 2002. V. **35**. P. L339), stimulates future research, in particular, the effects of cascade "evaporation" of electron shells of an atom (atomic ion) when filling an empty *K*-shell and can be tested in X-ray free-electron laser experiments with high peak radiation brightness and ultrashort photon pulse duration. The scheme of the proposed experiment with two linearly polarized X-ray photons is presented.

## 1. Introduction

The production of states with an *empty K*-shell of an atom [1,2] is one of the fundamental processes of the microcosm. Experimentally, it is identified through the registration $K^h_{\alpha,\beta}$ of hypersatellites of X-ray emission spectra and fine resonance structure of Auger- and photoelectron- spectra when an atom absorbs synchrotron radiation [3–11], radiation of an X-ray free-electron laser (XFEL) [12–15], and bombardment of an atom with fast heavy ions [16,17]. As far as we know, experimental studies of the generalized cross-section of the process of double ionization (sweeping) of the *K*-shell when an atom absorbs, for example, XFEL radiation of high peak brightness and ultrashort duration of the photon pulse [18–21] are still lacking. In the work of the authors [22], within the framework of the second-order nonrelativistic quantum perturbation theory and the Hartree-Fock single-configuration approximation, the first theoretical study of the generalized cross-section of the *direct* (through the birth of the *virtual* $1sxp(^1P_1)$– state of single ionization of the atom [23]) two-photon sweeping out of the *K*-shell of a light neon atom was carried out. At the same time, the effects of radial relaxation of transition states in the field of one and two $1s$–vacancies and only the projection $M = 0$ of the total angular momentum $J = 2$ of the wave functions of the finite $1s^0(l_1l_2; {}^1D_2)$–ionization states ($l_1l_1 = pp, sd$ – electrons of the continuous spectrum) were taken into account. In this paper, we modify the mathematical formalism of this work, take into account the non-trivial complete angular structure of the probability amplitudes of the transition to ${}^1D_2$–term ($J = 2, M = 0, \pm 1, \pm 2$) of the final states of double ionization, and generalize the theory to a heavy neon-like atomic ion. The object of the study is a neon-like ion of an iron atom (Fe$^{16+}$; the charge of the nucleus of the ion $Z = 26$; the configuration and term of the ground state $[0] = 1s^2 2s^2 2p^6 \; [{}^1S_0]$). The choice is due to the spherical symmetry of the ground state of the $Fe^{16+}$ ion and its availability in the gas phase [24] during experiments on the absorption of two linearly polarized XFEL photons of energy $\hbar\omega$ ($\hbar$ - Planck's constant, $\omega$ - circular frequency of the photon) by an ion trapped in the "trap" [25]. The spectral characteristics of the $Fe^{16+}$ ion are in high demand, in particular, in X-ray diagnostics of hot laboratory [26] and astrophysical [27] plasma, as well as structural materials in controlled thermonuclear fusion facilities [28]. It should be expected that the experimental detection of the effect of *direct* two-photon sweeping of the *K*-shell of an atom (atomic ion) at high energies of absorbed X-ray photons initiates, in particular, the generalization of the results of the study of the effect of cascade decay of one $1s$ - vacancy [29,30] on the case of a much more intense (the lifetime of an empty *K*-shell is ~ 2 times less than the lifetime of the $1s$–vacancies [31,32]) and structurally complex cascading filling of an empty *K*-shell.

## 2. Theory

Consider the following two-photon sweeping channels of the *K*-shell of a neon-like ion:

$$2\hbar\omega + [0] \to 1sxp + \hbar\omega \to 1s^0 yl_1 zl_2, \tag{1}$$

$$2\hbar\omega \geq I(1s^{-2}). \tag{2}$$

In (1), (2) the filled $2s^2$– and $2p^6$– shells a not specified, $x, y, z$ is the energies of electrons of the continuous spectrum, $l_1 l_2 = pp, ss, sd$ and $I(1s^{-2})$ is the energy of the threshold for the formation of an empty *K*-shell. The requirement to satisfy inequality (2) corresponds to the exclusion in (1) of the excited states of the discrete spectrum. Such consideration requires, first of all, the solution of the problem of taking into account the completeness of the set of intermediate and final states of a two-photon double excitation/ionization of an ion, and is the subject of future development of the theory. The structure of channels (1) corresponds to the following approximations. First. The strong spatial and energetic separation of the subvalent ($2s^2$) and valence ($2p^6$) shells from the deep $1s^2$ -shell of the $Fe^{16+}$ ion makes it possible to neglect the production of final $(1s2s, 1s2p^5, 2s^0, 2sp^5, 2p^4)ylzl'$ - ionization states. In fact, for the $Fe^{16+}$ ion, the inequalities are satisfied: $r_{1s}$ = 0.031 Å << $r_{2s}(r_{2p})$ = 0.140 (0.123) Å, $I_{1s}$ = 7699.23 eV >> $I_{2s}(I_{2p})$ = 1397.77 (1270.60) eV [33], where $r_{nl}$ ($I_{nl}$) is the average radius (ionization threshold energy) of the $nl$ -shell of the ion backbone. Second. The probability amplitude of two-photon *K*-shell sweeping through the channel $2\hbar\omega + [0] \to 1sxl \to 1s^0 yl_1 zl_2$ is determined by the operators of the contact [$\hat{C} \sim \sum_{n=1}^{N}(\hat{A}_n \cdot \hat{A}_n)$, $N$ is the number of electrons in the ion] and Coulomb interactions. As a result, it is proportional to the matrix element $\langle 1s | j_l | xl \rangle$, where $l$ is the symmetry of the electron of the virtual ionization state before its decay by the Coulomb interaction operator and $j_l$ is the spherical Bessel function are determined. For the electromagnetic field operator $\hat{A}_n$ (in the secondary quantization representation) the dipole approximation is assumed:

$$\hat{A}_n \to \sum_{\mathbf{k}} \sum_{\rho=1,2} \mathbf{e}_{\mathbf{k}\rho} (\hat{a}^+_{\mathbf{k}\rho} + \hat{a}^-_{\mathbf{k}\rho}), \tag{3}$$

$$(\mathbf{k} \cdot \mathbf{r}_n) << 1 \quad \Rightarrow \quad \exp[\pm i(\mathbf{k} \cdot \mathbf{r}_n)] \cong 1, \tag{4}$$

$$\theta_{1s} = \lambda_\omega / r_{1s} >> 1. \tag{5}$$

In (3–5) $\mathbf{e}_{\mathbf{k}\rho}$ ($\mathbf{k}$) is the polarization vector (wave vector) of the photon, $\hat{a}^+_{\mathbf{k}\rho}$ ($\hat{a}^-_{\mathbf{k}\rho}$) is the operator of the photon's production (annihilation), $\mathbf{r}_n$ is the radius-vector of the *n*- electron, and $\lambda_\omega$ is the wavelength of the absorbed photon. For the ion Fe$^{16+}$ at $\hbar\omega = 8$ keV we have $\theta_{1s} \cong 50$. Then $j_0 \to 1$, $j_{l\geq 1} \to 0$ and $\langle 1s|j_l|xl\rangle \to 0$. It should be noted here that in the recent work of the authors [34], the concept of the "applicability criterion" of the dipole approximation is modified when describing the photoabsorption cross-section of an atom. As a result, it is shown that the short-wave range of applicability of the dipole approximation (discarding, first of all, quadrupole corrections to the operator of the radiation transition) in terms of the energy of the absorbed X-ray photon is much wider than the region determined by the inequality (5).

The probability amplitudes of the two-photon sweeping of the *K*-shell of the ion through channels (1) are physically interpreted in Figs. 1a, b in the formalism of the (non-relativistic) Feynman diagrams. These diagrams are graphically disjointed. However, their analytic connectivity is encoded (through the insertion of Coulomb interaction lines) in the overlapping integrals of wave functions of single-particle states obtained in *different* Hartree-Fock fields [35]. The requirement for the coherence of Feynman diagrams, in particular, when describing the ground state energy of a system of many particles within the framework of nonrelativistic quantum perturbation theory is known as the *Goldstone-Hubbard theorem* on the expansion by connected diagrams [36].

Let us establish the analytical structure of the probability amplitudes of channel transitions (1). The following formulas of sections 2.1 – 2.4 of this work are given in atomic units (a.u.; $e = \hbar = m_e = 1$). In the calculation results (Fig. 1, 2, 3) and the **Appendix**, the usual units have been restored.

## 2.1. Amplitude along the $pp$–channel

According to Fig. 1a, for the desired amplitude with a fixed terms $T={}^1S_0$, ${}^1D_2$ of the final ionization state, we have:

$$A_M(T) = \sum_{M'} \int_0^\infty dx \frac{C_x \cdot \langle \Phi_x | \hat{R} | \Psi_{el} \rangle}{\varepsilon - x + i\gamma_{1s}}, \tag{6}$$

$$\hat{R} = -\frac{1}{c}\sum_{n=1}^N (\hat{p}_n \cdot \hat{A}_n), \tag{7}$$

$$|0\rangle = [0] \otimes (a_\omega^+)^2 |0_{ph}\rangle, \tag{8}$$

$$|\Phi_x\rangle = |1sxp({}^1P_1), M'\rangle \otimes \hat{a}_\omega^+ |0_{ph}\rangle, \tag{9}$$

$$|\Psi_{yz}\rangle = |1s^0 ypzp(T), M\rangle \otimes |0_{ph}\rangle. \tag{10}$$

In (6) – (10) the following are determined: $C_x = \langle 0|\hat{R}|\Phi_x\rangle$, the total wave functions of the initial ($|0\rangle$), intermediate ($|\Phi\rangle$) and final ($|\Psi\rangle$) states of two-photon ionization, $\hat{R}$ is the operator of the radiation transition, *c* is the speed of light in a vacuum, $\hat{p}_n$ is the operator of the momentum of the *n*-electron, the projections of the total angular momentum of the system "ionic residue ⊕ electrons" $M' = 0, \pm 1$, $M = 0$ for ${}^1S_0$-term, $M = 0, \pm 1, \pm 2$ for ${}^1D_2$–term, $|0_{ph}\rangle$ is the wave function of the photon vacuum of quantum electrodynamics, $\varepsilon = \omega - I_{1s}$ and $\gamma_{1s} = \Gamma_{1s}/2$, $\Gamma_{1s}$ is the width of the decay $1s$–vacancies of the virtual $1sxp$–state of ionization. Of course, the total wave functions (8)-(10) satisfy the general quantum-mechanical requirement of mutual orthogonality: $\langle 0|\Phi\rangle = \langle 0|\Psi\rangle = \langle \Phi|\Psi\rangle = 0$. Using the methods of algebra of photon production (annihilation) operators [37], the theory of irreducible tensor operators [38, 39], the theory of nonorthogonal orbitals [40] and taking an approximation for the integral of the overlapping radial parts of wave functions of single-electron states of a continuous spectrum,

$$\langle xp_+ | yp_{++}\rangle \cong \delta(x-y), \tag{11}$$

for amplitudes (6) we get:

$$A(^1S_0) = \frac{1}{\sqrt{3}} \cdot \xi \cdot \Lambda + (y \leftrightarrow z), \tag{12}$$

$$A_M(^1D_2) = \sqrt{5} \cdot \xi \cdot \Lambda \cdot Q_M + (y \leftrightarrow z), \tag{13}$$

$$\Lambda = \frac{(y+I_{1s})[z-I_{1s}+I(1s^{-2})]}{\varepsilon - y + i\gamma_{1s}} \langle 1s_0 \| \hat{r} \| yp_+ \rangle \langle 1s_+ | \hat{r} | zp_{++} \rangle, \tag{14}$$

$$\xi = \frac{4\pi}{3\omega V} \cdot N_{sp}, \quad N_{sp} = \langle 2s_+ | 2s_{++}\rangle^2 \langle 2p_+ | 2p_{++}\rangle^6, \tag{15}$$

$$Q_M = -\frac{4\pi}{3} \sum_{M'} \sum_p (-1)^{M'} Y_{1,M'}(\mathbf{e}_\omega) Y_{1,p}^*(\mathbf{e}_\omega) \begin{pmatrix} 1 & 1 & 2 \\ -M' & p & M \end{pmatrix}. \tag{16}$$

Here, through the permutation of $y-$ and $z-$ variables ($y \leftrightarrow z$), the quantum interference of amplitudes for fixed terms is taken into account and the following are determined: $\delta$ – Dirac delta function, $V(\text{cm}^3) = c$ is the volume of quantization of the electromagnetic field (numerically equal to the speed of light in a vacuum) [41], $Y_{\alpha,\beta}(\mathbf{e}_\omega)$ is the spherical function, $p = 0, \pm 1$, "*" is the symbol of complex conjugation, the 3$j$-Wigner symbol and the single-electron amplitude of the $1s \to yp$ transition probability:

$$\langle 1s_0 \| \hat{r} \| yp_+\rangle = N_{1s}(\langle 1s_0 | \hat{r} | yp_+\rangle - F), \tag{17}$$

$$N_{1s} = \langle 1s_0 | 1s_+\rangle \langle 2s_0 | 2s_+\rangle^2 \langle 2p_0 | 2p_+\rangle^6, \tag{18}$$

$$F = \frac{\langle 1s_0 | \hat{r} | 2p_+\rangle \langle 2p_0 | yp_+\rangle}{\langle 2p_0 | 2p_+\rangle}(-F), \tag{19}$$

$$\langle 1s_0 | \hat{r} | yp_+\rangle = \int_0^\infty P_{1s_0}(r) \cdot r \cdot P_{yp_+}(r) dr. \tag{20}$$

The indices "0, +, ++" correspond to the $P(r)$ – radial parts of the wave functions of electrons obtained by solving the self-consistent Hartree-Fock field equations for [0] –, $1s_+ yp_+$ – and $1s^0 yp_{++} zp_{++}$ – ion configurations.

In the **Appendix**, as an example, the mathematical details of constructing the amplitude of the $1sxp(^1P_1) \to 1s^0 ypzp(^1D_2)$ transition probability are given.

## 2.2. Amplitude along the $ss$–channel

According to Fig. 1b for the desired amplitude we have:

$$B(^1S_0) = \sum_{M'} \int_0^\infty dx \frac{C_x \cdot \langle \Phi_x | \hat{R} | S_{yz} \rangle}{\varepsilon - x + i\gamma_{1s}}, \tag{21}$$

$$|S_{yz}\rangle = |1s^0 yszs(^1S_0), M=0\rangle \otimes |0_{ph}\rangle. \tag{22}$$

Following the methods of constructing the amplitudes of section 2.1 and assuming the approximation of the substitution $x \to \varepsilon$ for the numerator of the subintegral expression in (21), we get:

$$B(^1S_0) \cong \xi \cdot \eta \cdot f(y) \cdot \langle \tau_\omega | \hat{r} | zs_{++} \rangle + (y \leftrightarrow z), \tag{23}$$

$$\eta = \omega \cdot [z + y - \omega + I(1s^{-2})]\langle 1s_0 \| \hat{r} \| \varepsilon p_+ \rangle. \tag{24}$$

In (23) $f(y) = \langle 1s_+ | ys_{++} \rangle$ and to take into account the *singular* single-electron matrix element of the operator of the radiation transition between *continuum* spectrum states ($\langle xp_+ | \hat{r} | zs_{++} \rangle \sim \sqrt{x}(x-z)^{-2} \to \infty$ at $x \to z$), a *correlation* function is determined:

$$|\tau_\omega(r)\rangle = \int_0^\infty dx \frac{|xp_+(r)\rangle}{\varepsilon - x + i\gamma_{1s}}. \tag{25}$$

The analytical result for (25) in the approximation of plane waves, $|xp(r)\rangle \sim \sin(kr)$, $k = \sqrt{2\varepsilon}$ is obtained in [42]:

$$|\tau_\omega(r)\rangle \cong -(2\pi/k)^{1/2} \cdot \exp[(ik - \gamma_{1s}/k) \cdot r]. \tag{26}$$

Taking into account (26) and the value of the improper integral of the first kind [43]

$$\int_0^\infty xe^{-px} \cdot \sin(bx) dx = \frac{2pb}{(p^2 + b^2)^2}, \tag{27}$$

in the approximation of the plane wave for the $|zs_{++}\rangle$–state, we get:

$$\langle \tau_\omega | \hat{r} | zs_{++} \rangle \cong -i\sqrt{2\varepsilon} \cdot \frac{1}{\gamma_{1s}^2}. \tag{28}$$

Thus, the appearance in (26) of the modulating function $\exp[-(\gamma_{1s}/k) \cdot r]$ with $\gamma_{1s} > 0$ leads to the *convergence* of integral (28) in force $|\tau_\omega(r)\rangle \to 0$ at $r \to \infty$.

Taking into account (28) for the desired amplitude, we get:

$$B(^1S_0) \cong \rho \cdot f(y) + (y \leftrightarrow z), \tag{29}$$

$$\rho = -i\xi\eta \cdot (\sqrt{2\varepsilon}/\gamma_{1s}^2). \tag{30}$$

## 2.3. Amplitude along the $sd$–channel

According to Fig. 1b for the desired amplitude we have:

$$B_M(^1D_2) = \sum_{M'} \int_0^\infty dx \frac{C_x \cdot \langle \Phi_x | \hat{R} | D_{yz} \rangle}{\varepsilon - x + i\gamma_{1s}}, \tag{31}$$

$$|D_{yz}\rangle = |1s^0 yszd(^1D_2), M\rangle \otimes |0_{ph}\rangle. \tag{32}$$

Following the methods of constructing the amplitudes of sect. 2.1, 2.2, for (31) we get:

$$B_M(^1D_2) \cong \sqrt{6} \cdot \rho \cdot f(y) \cdot Q_M + (y \leftrightarrow z). \tag{33}$$

## 2.4. Generalized ionization cross-section

Let us establish the analytical structure of the generalized cross-sections of the two-photon double ionization of an atomic ion [44, 45] along the $ll' = pp, ss, sd$ channels:

$$\sigma_g^{(ll')} = (V/2c) \int_0^\infty \int_0^\infty d^2\sigma^{(ll')}. \tag{34}$$

In (34), following Fermi's "*golden rule*" [46, 47], the usual double-differential cross-section of the process is determined:

$$d^2\sigma^{(ll')} = (\pi V/c) |M^{ll'}|^2 \cdot \delta(y + z - 2\omega + I(1s^{-2})) \, dydz, \tag{35}$$

$$|M_{pp}|^2 = |A(^1S_0)|^2 + \sum_M |A_M(^1D_2)|^2, \tag{36}$$

$$|M_{ss}|^2 = |B(^1S_0)|^2, \tag{37}$$

$$|M_{sd}|^2 = \sum_M |B_M(^1D_2)|^2. \tag{38}$$

Accounting for the lifetime of an empty $K$-shell ($\tau_{KK} \sim \gamma_{KK}^{-1}$, $\gamma_{KK} \cong 2\gamma_{1s}$ [31,32]) requires substitution in (35) $\delta(\Delta) \to (\gamma_{KK}/\pi)(\Delta^2 + \gamma_{KK}^2)^{-1}$, $\Delta = z + y - 2\omega + I(1s^{-2})$, and is the subject of future development of the theory.

When summing in (36) and (38) from the projections of the total moment of $^1D_2$ -term the final ionization states, the author's recent result [48,49] was taken into account:

$$\sum_M |Q_M|^2 = \frac{1}{5} - \frac{1}{20\pi}. \tag{39}$$

At the same time, the scheme of the *proposed* XFEL experiment for linearly polarized absorbed photons is implemented: $\mathbf{k} \in OZ$, $\mathbf{e}_\omega \in OX$ (*OX*, *OZ* are the axes of the rectangular coordinate system), $Y_{1,0}(\mathbf{e}_\omega) = 0$, $Y_{1,\pm 1}(\mathbf{e}_\omega) = \mp 3/(4\pi\sqrt{2})$.

When integrating into (34) for the $pp$ –channel in the numerator of the $\Lambda$–function from (14), the substitution approximation $y, z \to \varepsilon$ is assumed and the *Weisskopf-Wigner result* is taken into account [50]:

$$\int_{-\infty}^\infty \frac{dx}{(x^2 + \gamma_1^2)[(x - x_0)^2 + \gamma_2^2]} = \frac{\pi}{\gamma_1 \gamma_2} \frac{\gamma_1 + \gamma_2}{[x_0^2 + (\gamma_1 + \gamma_2)^2]}, \tag{40}$$

where in our case $x = y - \varepsilon$, $x_0 = 2I_{1s} - I(1s^{-2})$, $\gamma_1 = \gamma_2 = \gamma_{1s}$ and the lower limit of integration $x_{min} = 0$ is replaced by $x_{min} = -\infty$. When integrating in (34) for the $ss$– and $sd$ –channels for the $\eta$ –function in (24), the approximation of the substitution $y + z \to 2\omega - I(1s^{-2})$ is assumed [see $\delta$ –Dirac function in (35)].

Then, for the desired partial generalized cross-sections, we obtain (probabilities of photon disappearance without photoelectron registration):

$$\sigma_g^{(pp)} \cong \frac{2\pi}{\gamma_{1s}} \cdot \mu \cdot [\omega + I(1s^{-2}) - 2I_{1s}]^2 \cdot \left(\frac{4}{3} - \frac{1}{4\pi}\right) \cdot L^2, \tag{41}$$

$$\mu = \frac{8\pi^3}{9V} \cdot \alpha \cdot r_0^2 (a_0 c\hbar)^2 = 0.278 \cdot 10^{-52} \; [\text{cm}^4 \cdot \text{s}], \tag{42}$$

$$L = N_{sp} \left\langle 1s_0 \left\| \hat{r} \right\| \varepsilon p_+ \right\rangle \left\langle 1s_+ \left| \hat{r} \right| \varepsilon p_{++} \right\rangle, \tag{43}$$

$$\sigma_g^{(ss)} = \mu K^2 \cdot J, \tag{44}$$

$$K = N_{sp} \cdot \frac{\omega \sqrt{2\varepsilon}}{\gamma_{1s}^2} \left\langle 1s_0 \left\| \hat{r} \right\| \varepsilon p_+ \right\rangle, \tag{45}$$

$$J = \frac{a}{3} \cdot [8f^2(a/2) + (f(0) + f(a))^2], \tag{46}$$

$$\sigma_g^{(sd)} = \frac{6}{5}\left(1 - \frac{1}{4\pi}\right) \cdot \sigma_g^{(ss)}, \tag{47}$$

where $\alpha$ is the fine-structure constant, $r_0$ is the classical radius of the electron, $a_0$ is the Bohr radius and $a = 2\omega - I(1s^{-2})$. The result (47) shows a slight (~10%) difference between the $\sigma_g^{(ss)}$ – and $\sigma_g^{(sd)}$ – cross-sections.

## 3. Results and discussion

The calculation results are shown in Figs. 2,3. For the parameters of generalized cross-sections, the following values are taken: $\Gamma_{1s}$ = 1.046 eV [51], $I_{1s}$ = 7699.23 eV, $I(1s^{-2})$ = 15811.77 eV (relativistic calculation of this work) and $\omega \in (6; 100)$ keV.

According to Fig. 2, the following order of magnitude of the ratio of the generalized cross-sections of the two-photon sweeping of the K-shell is obtained:

$$(\sigma_g^{(ss)} + \sigma_g^{(sd)}) / \sigma_g^{(pp)} \cong 10^{13}. \tag{48}$$

The result (48) in the formalism of Feynman diagrams can be given the following physical interpretation. According to Fig. 1b, the first photon at a time $t_1$ ionizes the K-shell into a *virtual* $xp_+$ - state of the continuous spectrum. The second photon is absorbed at a moment of time $t_2 > t_1$ by the "cloud" of the $xp_+$- continuous spectrum. The $z(s,d)_{++}$ –continuum-spectrum state produced in this case by Coulomb repulsion ejects the remaining 1s-electron of the K-shell into the $ys_{++}$ –continuum state of the same symmetry (see overlapping integral $\left\langle 1s_+ | ys_{++} \right\rangle$). Such a process is much more likely than the process of free passage of the second photon through the "cloud" and its subsequent absorption by the remaining 1s - electron of the K-shell (Fig. 1a).

According to Fig. 3, for $2\omega \geq 2I_{1s}$ following order of magnitude of the ratio of the generalized cross-sections of two-photon sweeping and two-photon single ionization of the K-shell obtained:

$$(\sigma_g^{(ss)} + \sigma_g^{(sd)} + \sigma_g^{(pp)}) / (\sigma_g^{(s)} + \sigma_g^{(d)}) \cong 10^9. \tag{49}$$

The result (49) can be given the following physical interpretation. The probability of the $z(s,d)_{++}$ –continuum-spectrum state of ejecting the remaining $1s$ –electron of the K-shell into the $ys_{++}$– state of the continuous spectrum (Fig. 1b) is higher than the probability of its (1s–electron) not being "noticed" (see background section in Fig. 3).

Of course, the physical interpretation of the Feynman diagrams presented in Fig. 1a, b requires a rigorous mathematical justification. Such a justification requires the summation of an infinite functional series of connected diagrams on the basis of single-particle states obtained in a fixed, for example, a self-consistent Hartree-Fock field. Such a summation (if it is possible) is the subject of future development of the theory.

Finally, it should be noted that the maximum (at $\omega \cong 30$ keV) of the ratio of the generalized cross-sections of two-photon sweeping along the $pp$–channel (Fig. 1a) and two-photon single ionization (see extrapolation of the background cross-section in Fig. 3) of the $K$-shell was:

$$\sigma_g^{(pp)}/(\sigma_g^{(s)} + \sigma_g^{(d)}) \cong 6.5 \cdot 10^{-3}. \tag{50}$$

The result (50) is in order of magnitude consistent with the cross-section ratios of the single-photon double (Fig. 1c) and single (Fig. 1d) ionization of the $K$-shell ($\sim 10^{-3}$-$10^{-4}$) of an atom (atomic ion) known in the literature (see, e.g., [10,11]).

## 4. Conclusion

A nonrelativistic version of the quantum theory of the process of two-photon double ionization of the $K$-shell of a heavy neon-like atomic ion ($Fe^{16+}$) has been constructed. As the main result, it was found that in the region of energies of hard X-ray photons $2\omega \geq 15.8$ keV, the generalized cross-section of the two-photon sweeping of the $K$-shell significantly exceeds the generalized cross-section of the two-photon single ionization of the $K$-shell. The removal of the accepted approximations, in particular, going beyond the dipole approximation for the $\hat{R}$ ($\hat{C}$)–operator of the radiation (contact) transition and taking into account correlation and relativistic effects are the subject of future development of the theory. The generalization of the presented theory to atoms and atomic ions of other types and the establishment of the role of the charge of their nucleus is the subject of future research. Finally, the results of successful experiments on the observation of two-photon ionization of atoms, molecules, and solids (see [52-57] and references there) suggest that the absolute values of the generalized cross-section in Fig. 3 are quite measurable in the modern XFEL experiment.

**Appendix**

Moving on to the "radius form" for the $\hat{R}$ – operator from (7), given the matrix element of the photon production operator [58]

$$\langle \omega | \hat{a}^+_{\mathbf{k}\rho} | 0_{ph} \rangle = \left( \frac{2\pi c^2 \hbar}{\omega V} \right)^{1/2} \cdot \delta_{\mathbf{k}_\omega \mathbf{k}} \cdot \delta_{\rho_\omega \rho} \cong 10^9 \qquad (1A)$$

and the *Wigner-Eckart theorem* [59]

$$\langle 1s_+ xp_+; {}^1P_1, M' | Q^{(1)}_p | 1s^0 yp_{++} zp_{++}; {}^1D_2, M \rangle =$$
$$= (-1)^{1-M'} \cdot \Omega \cdot \begin{pmatrix} 1 & 1 & 2 \\ -M' & p & M \end{pmatrix}, \qquad (2A)$$

$$Q^{(1)}_p = \sqrt{4\pi/3} \cdot \sum_{n=1}^{N} Y_{1,p}(\mathbf{e}_n) \cdot r_n, \qquad (3A)$$

$$\Omega = (1s_+ xp_+; {}^1P_1 \| Q^{(1)} \| 1s^0 yp_{++} zp_{++}; {}^1D_2), \qquad (4A)$$

for the amplitude (6) of the probability of $1sxp({}^1P_1) \to 1s^0 ypzp({}^1D_2)$ transition we have:

$$\langle \Phi_x | \hat{R} | \Psi_{yz} \rangle = \chi \Omega \cdot \varphi_{MM'}, \qquad (5A)$$

$$\chi = -i \left( \frac{2\pi e^6}{\hbar \omega V} \right)^{1/2} (x - y - z + I_{1s} - I(1s^{-2})), \qquad (6A)$$

$$\varphi_{MM'} = -\sqrt{4\pi/3} \cdot \sum_p (-1)^{p-M'} \cdot Y_{1,-p}(\mathbf{e}_\omega) \cdot \begin{pmatrix} 1 & 1 & 2 \\ -M' & p & M \end{pmatrix}, \qquad (7A)$$

where $\mathbf{k}_\omega$ ($\rho_\omega$) is the wave vector (polarization) of the absorbed photon, $\mathbf{e}_n = \mathbf{r}_n / r_n$ and $r_n$ is the length of the vector $\mathbf{r}_n$.

Let us take into account the result of the monograph [38] (*Chapter* 3, § 29) for the reduced matrix element of the general form:

$$(n_0 l_0^{N_0} n_1 l_1^{N_1} n_2 l_2^{N_2} T_{01} T_2, LSJ \| Q^{(k)} \| n_0 l_0^{N_0} n_1 l_1^{N_1-1} n_2 l_2^{N_2+1} T'_{01} T'_2, L'S'J'), \qquad (8A)$$

where $T_{01} = L_{01} S_{01}$ is the resulting term of $n_0 l_0$ – and $n_1 l_1$ –shells, $T_2 = L_2 S_2$ is the resulting term of $n_2 l_2$ -shell, $LSJ$ is the term of the whole state, and the $n_0 l_0$ - shell plays the role of an "*observer*". In our case, $l_0 = l_2 = 1$, $l_1 = 0$, $N_0 = N_1 = 1$, $N_2 = 0$, $T_{01} = {}^1P$, $T_2 = {}^1S$, $LSJ = {}^1P_1$, $T'_{01} = T'_2 = {}^2P$, $L'S'J' = {}^1D_2$, genealogical coefficients $(s({}^1P) \| s^0({}^2P)s) = (p^0({}^1S)p \| p({}^2P)) = 1$ and [60]

$$\begin{Bmatrix} 1 & 1 & 0 \\ 2 & 2 & 1 \end{Bmatrix} = \frac{1}{\sqrt{15}}, \quad \begin{Bmatrix} 0 & 0 & 0 \\ 1/2 & 1/2 & 1/2 \end{Bmatrix} = -\frac{1}{\sqrt{2}}, \quad \begin{Bmatrix} 1 & 1 & 0 \\ 0 & 1 & 1 \\ 1 & 2 & 1 \end{Bmatrix} = \frac{1}{9}. \qquad (9A)$$

Then from (8A) for $k = 1$ we get:

$$\Omega = \sqrt{5/3} \cdot \langle 1s_+ | \hat{r} | zp_{++} \rangle \langle xp_+ | yp_{++} \rangle \cdot N_{sp}, \qquad (10A)$$

which completes the construction of the amplitude (5A).

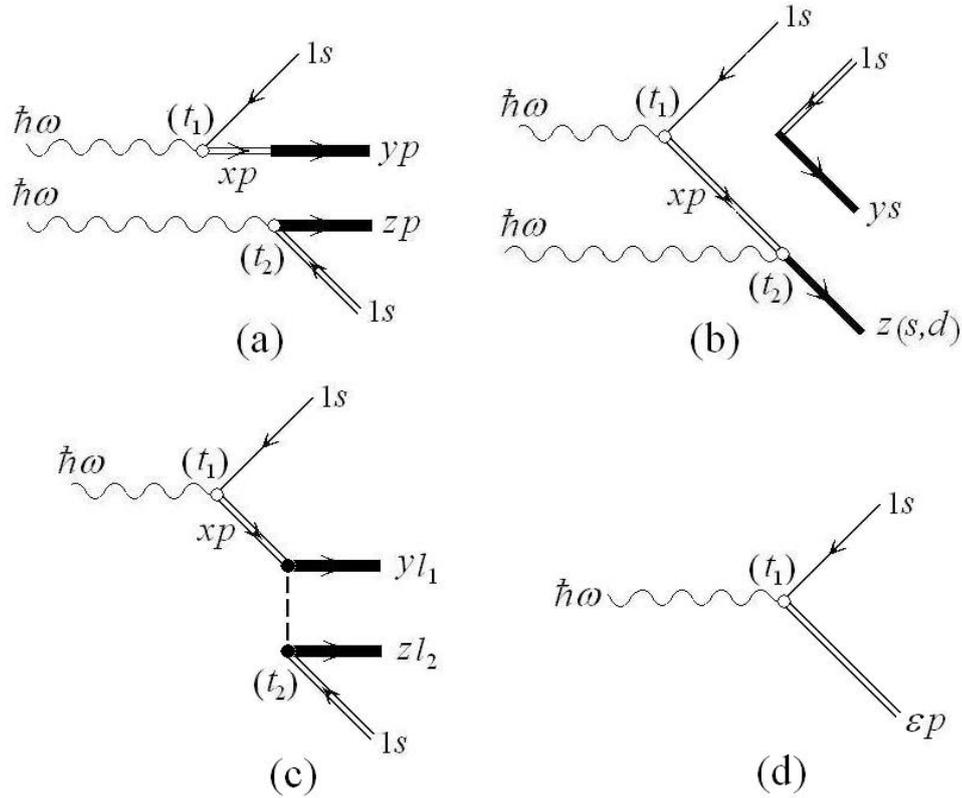

**Figure 1.** Probability amplitude of two-photon and single-photon ionization of the *K*-shell of a neon-like atomic ion ($Fe^{16+}$) in the representation of (non-relativistic) Feynman diagrams: (a) on channel (1) for $l_1 l_2 = pp$; (b) by channels (1) for $l_1 l_2 = ss, sd$; (c) over the channels $\hbar\omega + [0] \to 1sxp \to 1s^0 yl_1 zl_2$ for $l_1 l_2 = sp, pd, df, \ldots$ ; (d) on the channel $\hbar\omega + [0] \to 1s\varepsilon p$, $\varepsilon = \hbar\omega - I_{1s}$. An arrow to the left is a vacancy, an arrow to the right is an electron. Double (wide black) line – the states is obtained in the Hartree-Fock field of one (two) $1s$-vacancies. The junction of the double and wide black lines corresponds to the overlapping integral ($\langle xp | yp \rangle, \langle 1s | ys \rangle$). Dotted line – Coulomb interaction. A light circle is the top of the radiation transition. The direction of time is from left to right ($t_1 < t_2$). $\hbar\omega$ is the energy of the absorbed photon.

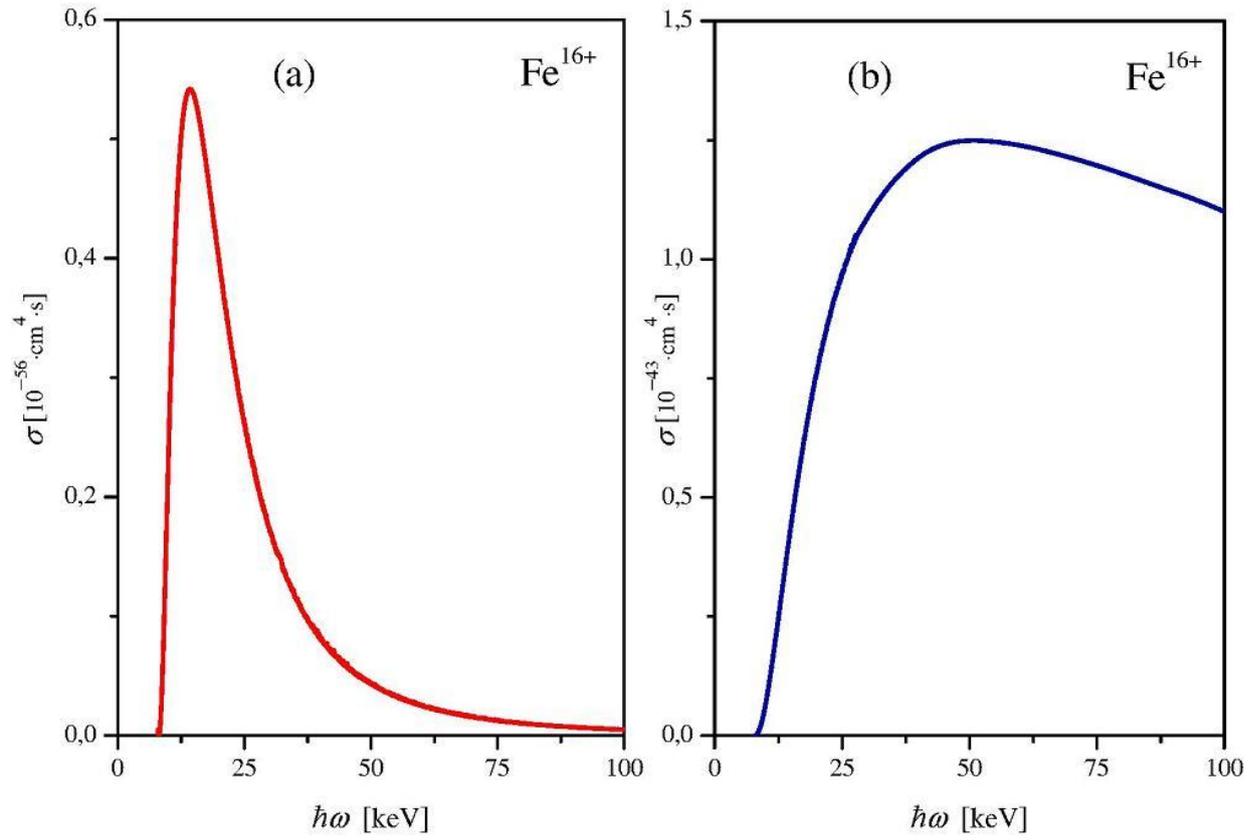

**Figure 2.** Partial generalized cross-sections of the two-photon double ionization of the *K*-shell of the Fe$^{16+}$ ion: (a) along the channel (1) for $l_1 l_2 = pp$ (see Fig. 1a); (b) by channels (1) for $l_1 l_2 = ss, sd$ (see Fig. 1b). $\hbar\omega$ is the energy of the absorbed photon.

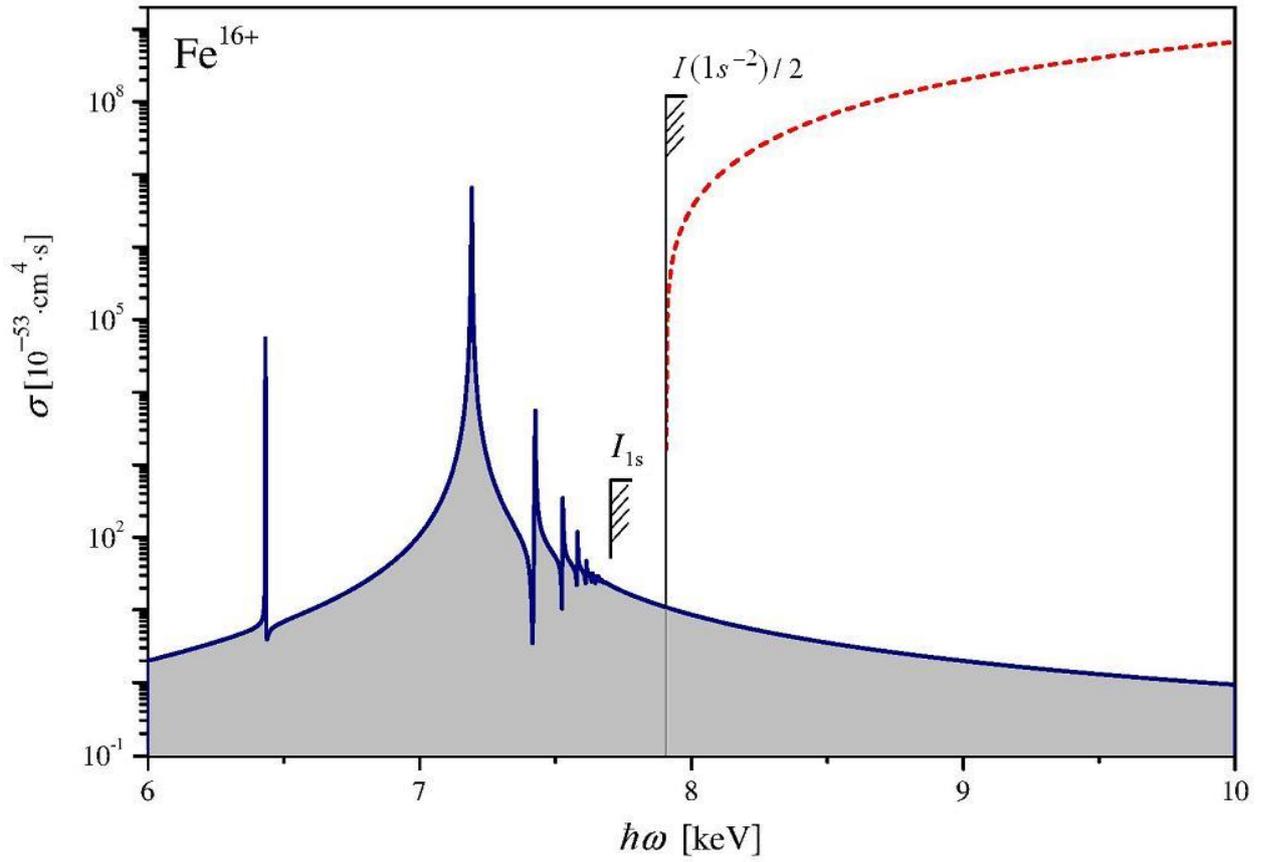

**Figure 3.** The total (by channels (1); Fig. 2) generalized cross-section of the two-photon double ionization of the *K*-shell of the Fe$^{16+}$ ion (dotted curve for $\hbar\omega \geq I(1s^{-2})/2$). A generalized cross-section of a two-photon single ionization of the Fe$^{16+}$ ion of the work [48] for $\hbar\omega \in (6; 10)$ keV is given as a background. $\hbar\omega$ is the energy of the absorbed photon.